\documentclass[conference,a4paper]{IEEEtran}

\usepackage{paralist}
\usepackage{enumitem}
\usepackage{cite}
\usepackage[hyphens]{url}
\usepackage{xcolor}
\usepackage[printonlyused,nolist]{acronym}
\usepackage{soul}
\usepackage{graphicx}

\makeatletter
\def\ps@IEEEtitlepagestyle{%
\def\@oddfoot{\parbox{\textwidth}{\footnotesize
This is the author's version of a paper accepted for publication in the proceedings of the 2020 6th IEEE International Energy Conference (ENERGYCon). 
It is posted here by permission of IEEE. 
Not for redistribution. 
Cite as:
D.~van der Velde \emph{et~al.}, ``{Methods for Actors in the Electric Power System to Prevent, Detect and React to ICT Attacks and Failures},'' in \emph{Proceedings of the 2020 6th IEEE International Energy Conference (ENERGYCon)}, 2020.
\\[1em]
\textcopyright{} 2020 IEEE. 
Personal use of this material is permitted.  
Permission from IEEE must be obtained for all other uses, in any current or future media, including reprinting/republishing this material for advertising or promotional purposes, creating new collective works, for resale or redistribution to servers or lists, or reuse of any copyrighted component of this work in other works.\vspace{3em}}
}%
}
\makeatother

\begin{document}

\title{\fontdimen2\font=6.4pt Methods for Actors in the Electric Power System to Prevent, Detect and React to ICT Attacks and Failures}

\author{
\IEEEauthorblockN{%
Dennis van der Velde\IEEEauthorrefmark{3},
Martin Henze\IEEEauthorrefmark{1},
Philipp Kathmann\IEEEauthorrefmark{2},
Erik Wassermann\IEEEauthorrefmark{4},\\
Michael Andres\IEEEauthorrefmark{3},
Detert Bracht\IEEEauthorrefmark{8},
Raphael Ernst\IEEEauthorrefmark{1},
George Hallak\IEEEauthorrefmark{6},
Benedikt Klaer\IEEEauthorrefmark{3},\\
Philipp Linnartz\IEEEauthorrefmark{5},
Benjamin Meyer\IEEEauthorrefmark{7},
Simon Ofner\IEEEauthorrefmark{1},
Tobias Pletzer\IEEEauthorrefmark{4},
Richard Sethmann\IEEEauthorrefmark{2}
}
\IEEEauthorblockA{%
\IEEEauthorrefmark{6}\textit{devolo AG,} Aachen, Germany $\cdot$ george.hallak@devolo.de}

\IEEEauthorblockA{%
\IEEEauthorrefmark{3}\textit{Fraunhofer FIT,} Aachen, Germany $\cdot$ \{dennis.van.der.velde, michael.andres, benedikt.klaer\}@fit.fraunhofer.de}

\IEEEauthorblockA{%
\IEEEauthorrefmark{1}\textit{Fraunhofer FKIE,} Wachtberg, Germany $\cdot$ \{martin.henze, raphael.ernst, simon.ofner\}@fkie.fraunhofer.de}

\IEEEauthorblockA{%
\IEEEauthorrefmark{7}\textit{KISTERS AG,} Aachen, Germany $\cdot$ benjamin.meyer@kisters.de}

\IEEEauthorblockA{%
\IEEEauthorrefmark{5}\textit{RWTH Aachen University,} Aachen, Germany $\cdot$ p.linnartz@iaew.rwth-aachen.de}

\IEEEauthorblockA{%
\IEEEauthorrefmark{4}\textit{Schleswig-Holstein Netz AG,} Quickborn, Germany $\cdot$ \{erik.wassermann, tobias.pletzer\}@sh-netz.com}

\IEEEauthorblockA{%
\IEEEauthorrefmark{8}\textit{umlaut energy GmbH,} Aachen, Germany $\cdot$ detert.bracht@umlaut.com}

\IEEEauthorblockA{%
\IEEEauthorrefmark{2}\textit{University of Applied Sciences Bremen,} Bremen, Germany $\cdot$ \{philipp.kathmann, richard.sethmann\}@hs-bremen.de}

}

\maketitle

\begin{abstract}
The fundamental changes in power supply and increasing decentralization require more active grid operation and an increased integration of ICT at all power system actors.
This trend raises complexity and increasingly leads to interactions between primary grid operation and ICT as well as different power system actors.
For example, virtual power plants control various assets in the distribution grid via ICT to jointly market existing flexibilities.
Failures of ICT or targeted attacks can thus have serious effects on security of supply and system stability.
This paper presents a holistic approach to providing methods specifically for actors in the power system for prevention, detection, and reaction to ICT attacks and failures.
The focus of our measures are solutions for ICT monitoring, systems for the detection of ICT attacks and intrusions in the process network, and the provision of actionable guidelines as well as a practice environment for the response to potential ICT security incidents.
\end{abstract}


\begin{acronym}
\acro{BMWi}{German Federal Ministry for Economic Affairs and Energy}
\acro{DSO}{Distribution System Operator}
\acro{ICT}{Information and Communication Technology}
\acro{IDS}{Intrusion Detection System}
\acro{OT}{Operational Technology}
\acro{RTU}{Remote Terminal Unit}
\acro{SCADA}{Supervisory Control and Data Acquisition}
\acro{SMGW}{Smart Meter Gateway}
\acro{TSO}{Transmission System Operator}
\acro{VPP}{Virtual Power Plant}
\end{acronym}

\section{Introduction}

Power supply is undergoing a fundamental change as \ac{ICT} systems are implemented by all power system actors~\cite{panajotovic_ict_2011}.
At the same time, the increased expansion of distributed generation and the integration of new consumer types (e.g., charging infrastructure) at the distribution grid level, require rethinking conventional grid operation~\cite{santodomingo_sgam_2014,mckeever_bottom_2016}.
Different actors, grid operators, energy production, as well as private and industrial customers increasingly interact, e.g., by offering flexibility and market mechanisms controlled and enabled by \ac{ICT} components.

As a result, malfunctions on the \ac{ICT} level can directly impact physical grid operation, potentially leading to local or large-scale power outages.
This makes the power system \ac{ICT} infrastructure---\acp{TSO}, \acp{DSO}, \ac{VPP} operators, metering point operators, and manufacturers---an attractive target for advanced cyber attacks~\cite{skopik_dealing_2014}, as indicated by the successful attacks on the \ac{ICT} infrastructure of the Ukrainian power grid in 2015 and 2016~\cite{lee_tlp_2016,lee_crashoverride_2017}.

To prevent such cyber attacks, existing security measures must be accompanied by a close monitoring of irregularities and fast incident response.
This requires a comprehensive \ac{IDS} as a measure for \ac{ICT} networks of power system actors that takes domain-specific characteristics (e.g., process information) into account~\cite{genzel_security_2017}.
To obviate the need for long pilot phases for integration into grid operation, current and advanced security technologies need to be tested in interaction with primary technology in a realistic environment.
Finally, as even the best security measures do not offer full protection against complex attacks, non-cyber-security experts need technical guidelines to quickly decide if a given event is a cyber attack or failure.

Only an optimal interplay of prevention, detection, and reaction allows to counter new advanced and unknown threats to \ac{ICT} infrastructure in power systems and thus contribute to system safety and availability.
To achieve this goal, we present our ongoing work on MEDIT to provide \textbf{m}ethods for actors in the \textbf{e}lectric power system to prevent, \textbf{d}etect and react to \textbf{I}C\textbf{T} attacks and failures along the following three segments:

\begin{enumerate}[nosep,leftmargin=1.2em]
\item As basis for our work, a \emph{research and validation environment} for future distribution grids allows to investigate complex interactions or deviations from normal operation, e.g., through \ac{ICT} attacks and failures. 
Here, MEDIT relies on an \ac{ICT} energy co-simulation environment for large-scale investigation and a realistic distribution grid laboratory. 

\item MEDIT's approach for the \emph{prevention and detection of \ac{ICT} failures and attacks} on uses monitoring for power system components to record and observe parameters of technical systems to detect abnormal behavior as well as intrusion detections systems to timely detect the active intrusion of third parties into the \ac{ICT} infrastructure of energy actors.

\item To support the \emph{reaction to \ac{ICT} security incidents}, MEDIT relies on a realistic training and simulation environment to teach technical personnel as well as actionable incident response guidelines that allow non-security experts to adequately mitigate and repel \ac{ICT} failures and attacks.
\end{enumerate}

\section{\ac{ICT} Infrastructure and Attack Vectors in Distribution Grids}
\label{sec:background}

The demand for real-time information and control of distributed assets for new operational concepts~\cite{santodomingo_sgam_2014,mckeever_bottom_2016}, requires an extensive and partly parallel \ac{ICT} infrastructure via private or public communication channels.
This creates new risks and attack vectors for the entire energy supply system~\cite{skopik_dealing_2014,genzel_security_2017}.

In Figure~\ref{fig:scenario}, we exemplary give an overview of the envisioned \ac{ICT} infrastructure of a \ac{DSO} in future power systems.
The central \ac{SCADA} system, which controls the distribution grid, is connected through firewalls to multiple networks such as the company network and the Internet, often used for remote access.
To ensure security for this remote access, an additional terminal server separates connections and implements additional security mechanisms such as two factor authentication.

For the actual grid operation, however, the \ac{DSO} uses a completely private \ac{ICT} infrastructure owned and operated by the \ac{DSO}.
This communication network is often called \ac{OT} network and is separated from the normal company network, used for the conventional office IT such as desktop PCs and printers.
Through this \ac{OT} network, the \ac{SCADA} system receives measurements from various grid assets, e.g., voltage and currents from transformers in substations, and can even remotely control parts of these assets.

In the past, such control was usually limited to substations or on-load tap changing transformers.
With the increased expansion of distributed generation units and the integration of new types of consumers, more and more assets are directly connected to the \ac{SCADA} system.
Examples include distributed energy resources such as wind turbines or photovoltaic systems and intelligent new measurement systems such as \acp{SMGW} which may provide new sensory input for the \ac{SCADA} systems, as well as new controllable consumers such as electrical vehicle charging stations or battery systems.

Based on this \ac{ICT} infrastructure, we can deduce typical communication patterns and group them into typical use cases for different actors in the power system.
Using the resulting use cases and communication patterns, we identify the following clusters of likely attack vectors:

\begin{figure}
\centering
\includegraphics{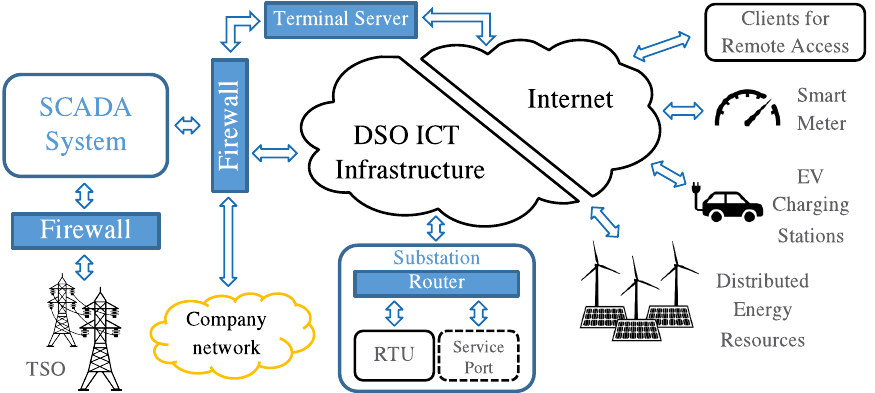} 
\caption{In a future \ac{ICT} infrastructure of a \ac{DSO}, grid assets are connected to the \ac{SCADA} system through dedicated infrastructure and\slash or the Internet.}
\label{fig:scenario}
\vspace{-1em}
\end{figure}

\noindent\textbf{Direct attack on \ac{SCADA} system:}
An attacker attempts to directly manipulate the \ac{SCADA} system itself to gain control over it, disable it, or provoke its malfunction.

\noindent\textbf{Indirect attack on \ac{SCADA} system:}
An attacker attempts to manipulate information to provoke wrong decisions in the \ac{SCADA} system (e.g., using false data injection).

\noindent\textbf{Attack on \ac{OT} network:}
An attacker attempts to manipulate or impede functionalities of active \ac{OT} network components (e.g., switches\slash routers), leading to malfunction or unavailability.

\noindent\textbf{Attack on \acp{RTU} and attached devices:}
An attacker attempts to manipulate \acp{RTU} to gain control over them, disable them, or provoke their malfunction.

Importantly, the inclusion of more and more intelligent assets into the \ac{ICT} infrastructure results in the increase of possible attack vectors.
In general, each interface to each asset is a potential attack vector, which could be used to compromise the \ac{OT} network or the \ac{SCADA} system.
Furthermore, each asset for itself is a potential target, either to gain control over it, disable it or use it as a bridge-head for further attacks.

Our analysis of attack vectors in the \ac{ICT} infrastructure of \acp{DSO} highlights the imperative need to support actors in the power system with methods to detect, prevent, and react to \ac{ICT} attacks and failures.
While certain attack vectors exhibit only little specifics to the energy sector and can thus be countered with traditional \ac{ICT} security mechanisms~\cite{henze_cps-security_2017,roepert_opcua_2020}, a plethora of threats and risks specific to power systems needs to be accounted for.
Our ongoing work on MEDIT targets specifically these threats and risks, striving for solutions that can practically be deployed in the \ac{ICT} infrastructure of \acp{DSO}.

\section{Research and Validation Environment}
\label{sec:research-and-validation-environment}

As MEDIT aims at the development of practically usable and deployable \ac{ICT} security methods specifically for the power system domain, we require both an appropriate simulation environment and a distribution grid laboratory.

\subsection{\ac{ICT} Energy Co-Simulation Environment}
\label{sec:co-simulation}

Since both the behavior of the power system and the associated \ac{ICT} network are relevant to develop and evaluate \ac{ICT} security technology, an appropriate simulation environment has to map both domains and their interactions in a suitable level of detail.
Complex interactions between \ac{ICT} and the grid should be considered to examine and evaluate questions regarding the requirements of \ac{ICT} used and repercussions on security of supply and system stability through \ac{ICT} attacks or failures in a scalable and systematic manner~\cite{mueller_cosim_2016}.
To simulate attacks on \ac{ICT}, special requirements are placed on the \ac{ICT} simulation site.
The main requirement is an explicit and extendable simulation of secondary equipment and their protocol-compliant communication as realistic as necessary.
Furthermore, techniques used in \ac{ICT} networks such as routing and packet switching must be simulated realistically to enable authentic attack simulation.
In particular, the environment must enable the generation of extensive training data for an \ac{IDS}.
MEDIT evaluates different \ac{ICT} simulators for their suitability.
Only through a realistic simulation, we can develop and train an \ac{IDS} before validating it in a laboratory and finally deploying it in a real environment.

Co-simulation is a valid approach to couple simulators from different domains and model their interdependence.
Different co-simulation standards, e.g., HLA~\cite{IEEE_HLA_2010} and FMI~\cite{blochwitz_fmi_2012}, as well as more specific frameworks, e.g., MOSAIK~\cite{steinbrink_mosaik_2019} and Ptolemy II~\cite{ptolemaeus_ptolemy_2014}, are available.
Especially for the simulation of smart grids, co-simulation has already been investigated from different perspectives~\cite{steinbrink_cosimulation_2018,vogt_cosimulation_2018}.
For example, SGsim~\cite{awad_sgsim_2014} uses OpenDSS for the power flow simulation and combines it with OMNeT++~\cite{varga_omnet_2001} to simulate the \ac{ICT} network.

While these standards and preliminary works provide valuable input for our work, we require a co-simulation that allows us to integrate the developed \ac{ICT} monitoring and \acp{IDS} into the simulation environment and a protocol-compliant communication between components on the \ac{ICT} site.
Consequently, we use the approach of co-simulation to couple suitable, independent simulation environments for our purposes.

\begin{figure}
\centering
\includegraphics[width=0.9\columnwidth]{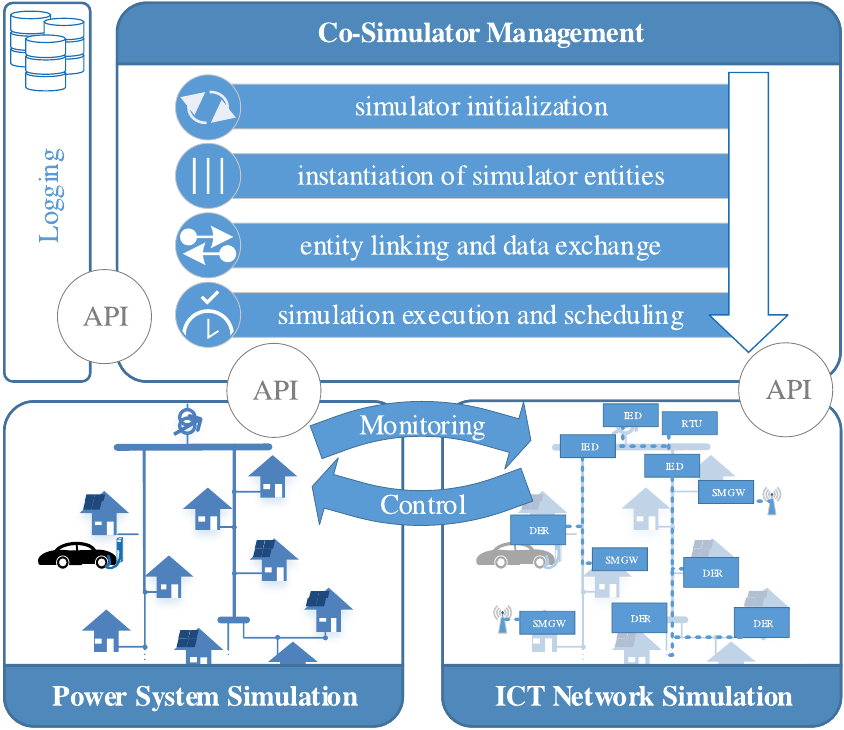} 
\caption{MEDIT's co-simulation environment uses a co-simulation manager to interface with domain specific simulators for power systems and \ac{ICT}.}
\label{fig:cosim}
\vspace{-1em}
\end{figure}

The concept of our environment is shown in Figure~\ref{fig:cosim}.
To create uniform interfaces between the domain specific simulators, MOSAIK~\cite{steinbrink_mosaik_2019} will be used as a co-simulation manager.
For the simulation of power systems, static simulations based on PANDAPOWER~\cite{thurner_pandapower_2018} as well as dynamic simulations for examination of power system stability based on MATPAT~\cite{roehder_matpat_2017} are integrated into the co-simulation environment.
Here, we will focus on detailed modeling of medium and subordinate low-voltage grids with a high number of distributed resources.

Based on primary technology used in the power grid, communication processes and technologies as well as generated data flows are modeled via standardized protocols such as IEC 60870-5-104 and IEC 61850 to generate real process data traffic.
The environment functionally maps the complete communication chain between the operational layer, station layer, and the field layer for \acp{DSO}.
This includes modeling \ac{RTU} functionality for aggregation and protocol conversion of process data and modeling of intelligent electronic device functionality as links to primary technology, locally implementing control and monitoring capability.
This at the same time represents the logical interface between power grid simulation and network simulation.
In addition, we also cover the influence of other actors on grid operation, e.g., by \acp{VPP} operators through the simultaneous control of grid resources.

By mapping real process data traffic in the simulation, it will also be possible to link the simulation with real control systems and real secondary equipment of the actors under consideration within our distribution grid laboratory.
Thus, the simulation can be used for scaling the emerging process data traffic within a real environment as described in the following.

\subsection{Distribution Grid Laboratory}
\label{sec:laboratory}

Even if the simulation thus approximates reality, the integration of the developed technologies in a real laboratory for validation purposes is used to evaluate the integrability into real power systems.
In particular, more complex systems such as a \ac{SMGW} infrastructure cannot be completely represented in a simulation in accordance with reality and are therefore set up in the laboratory environment, which is utilized as one essential part of MEDIT's research and validation environment.

\begin{figure}
\centering
\includegraphics{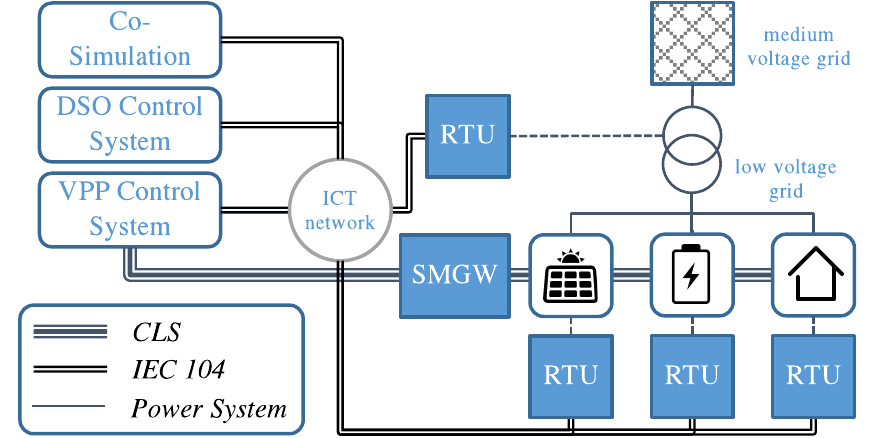} 
\caption{A 10 kV distribution grid laboratory is extended with \ac{ICT} infrastructure to represent the whole cyber-physical system of distribution grids.}
\label{fig:lab}
\vspace{-1em}
\end{figure}

The laboratory at RWTH Aachen University consists of a medium/low voltage distribution grid including four distribution substations (10\,kV\slash 400\,V), several distributed generation units (e.g., battery storage systems (15\,kVA-100\,kVA), photovoltaic inverters (10\,kVA-36\,kVA), and an 50\,kW combined heat and power emulator), loads, secondary equipment and measurement devices, that can be interconnected flexibly.
As shown in Figure~\ref{fig:lab}, our existing laboratory is extended by the required \ac{ICT} infrastructure to depict the whole cyber-physical system of the distribution grid, including \ac{ICT} and relevant processes of the different power system actors (e.g., \ac{DSO} and \ac{VPP} operators).
Our focus is on providing an infrastructure as close as possible to the reality.
Therefore, the laboratory extension is conducted in accordance with the requirements of \acp{DSO} and \acp{VPP} operators.

Via secondary equipment and \ac{ICT}, the assets are connected to a local distribution grid control system using the IEC 60870-5-104 protocol.
Additionally, they are also connected to a \ac{VPP} control system provided in the cloud using the IEC 60870-5-104 protocol and the Controllable Local System (CLS) interface of the \acp{SMGW}.
Suitable data points are defined for each asset and thus realistic process data traffic (control signals, measuring signals, status signals) will be generated, sent, and processed.
Since the laboratory cannot provide an unlimited number of (physical) assets, it will be linked with the co-simulation environment in the ICT-domain to enable the generation of artificial yet validated data traffic to be sent to the control systems to demonstrate scalability.
Consequently, the distribution grid laboratory is used to show the impact of cyber attacks on the \ac{ICT} and on primary equipment as well as to validate the artificially generated data traffic and the developed tools for prevention, detection, and reaction.

\section{Prevent and Detect \ac{ICT} Failures and Attacks}
\label{sec:prevention}

MEDIT's methods to prevent and detect \ac{ICT} failures and attacks particularly focus on suitable solutions for \ac{ICT} monitoring and intrusion detection for power system actors.

\subsection{\ac{ICT} Monitoring System for Distribution Grids}

\ac{ICT} monitoring primarily serves to record and monitor parameters and status of technical systems.
This makes it possible to detect, visualize, and react to malfunctions based on data from different sources.
In power systems, malfunctions and disturbances can increasingly no longer be attributed to primary technology alone.
Indeed, current events indicate that errors may increasingly occur on the IT and \ac{OT} side~\cite{lee_tlp_2016}.
As corresponding monitoring at \ac{ICT} level is not yet fully implemented in practice, there is a risk that \ac{ICT}-caused malfunction cannot be adequately attributed, leading to a ``blind spot''~\cite{whitepaper}.
To ensure reliable fault diagnosis, complex on-site procedures are therefore required.
This can endanger network availability and system security, resulting in higher downtime~\cite{genzelmoderne}.

To address this issue, MEDIT will develop an \ac{ICT} monitoring system that identifies \ac{ICT} error sources.
Currently, \acp{DSO} use different suppliers for \ac{ICT} components to ensure independent and reliable grid operation.
Ensuring independence from manufacturers is reasonable, but complicates data collection for monitoring.
Consequently, MEDIT has to evaluate existing monitoring protocols for their applicability.
Based on this, a suitable monitoring solution has to be implemented and integrated.
This involves working with existing data sources or data sources that can be made available with little effort.

The monitoring system should serve as a troubleshooting reference point for control center personnel without deeper \ac{ICT} expertise.
Thus, personnel can recognize whether a disturbance of the \ac{ICT} infrastructure or of primary equipment is present.
As a more advanced concept, security monitoring could trace the current status of devices more closely with regard to \ac{ICT} security.
Technologies such as hardware security modules can be used to better measure the security state.
Our approach provides a solution for the current need, but also shows the advantages of extended monitoring.
Power system actors can thus find a sweet spot of costs and benefits in terms of monitoring and continuously adjust their strategy.

\subsection{Intrusion Detection System for Electrical Process Data}
\label{sec:intrusion-detection}

Coordinated \ac{ICT} attacks on power system actors have a high threat potential.
In \ac{ICT} networks, using \acp{IDS} is a state of the art method to timely detect the active intrusion of third parties into the \ac{ICT} infrastructure and thus enable countermeasures~\cite{gorzelak2011proactive}.
\acp{IDS} can be classified on the basis of used detection methods (signature\slash anomaly\slash specification-based), deployments (host\slash network), used information (network\slash protocol\slash process data), and inclusion of history (stateless\slash state-oriented)~\cite{morin2007intrusion}.
In contrast to office IT, process networks in the power system domain have a less heterogeneous structure.
This results in a comparatively low number of different device types and applications and longer life span of devices, which allows MEDIT to specifically focus on power system specific communication protocols such as IEC 60870-5-104.
Endpoints in these \ac{ICT} networks fulfill a clearly defined purpose and are therefore limited to a known set of commands as well as communicate only with explicitly and upfront determined entities, which eases the specification of network compliant behavior~\cite{serror_innetwork_2018}.

To detect intrusions on the protocol layer, MEDIT will improve upon the protocol whitelisting approach~\cite{yang2014stateful}, a three-step approach following the principle of security in depth based on
\begin{inparaenum}[(i)]
	\item access-control detection,
	\item protocol whitelisting detection, and
	\item model-based detection.
\end{inparaenum}
Within the \ac{ICT}, we require network borders that define physical units, e.g., substations.
Then, the goal of protocol whitelisting is to detect attacks that
\begin{inparaenum}[(i)]
\item connect to a substation network,
\item send or manipulate \ac{RTU} traffic, or
\item scan the network.
\end{inparaenum}

For access-control detection, our \ac{IDS} needs to monitor existing network participants on the OSI Layers 2-4.
Protocol whitelisting detection monitors which protocols are used between endpoints in the network, while a deeper protocol analysis is performed by model-based detection.
To perform protocol whitelisting, the \ac{IDS} needs to know the network configuration and allowed communication patterns.
MEDIT will propose a solution for the \ac{IDS} whitelisting approach for power system specific protocols ~\cite{yang2014stateful}.

Furthermore, we will investigate to which extent anomalies in the area of grid operation can be evaluated as indicators of successful attacks on compromised areas of secondary and control technology.
Examples are switching commands or setpoint adjustments that have led to a traceable blackout.
Faulty measured values or messages identified by bad data detection as well as implausible load, generation and substation behavior and limit violations could also serve as indicators~\cite{constante_mc-attacks_2019}.
Attacks on the parametrization and integrity of the process data are therefore mapped in advance in the co-simulation and detected by a distributed network-based intrusion detection system.
On the basis of stochastic methods and graph-based machine learning methods, e.g., GraphSAGE~\cite{hamilton_graphsage_2018}, these indicators and alarms from signature-based and whitelist methods can be correlated and the attack scenario can be reconstructed topologically and chronologically.
The result of this correlation is an assessment of the state of compromising for individual \ac{OT} and \ac{ICT} devices as decision support for business continuity.

\section{React to \ac{ICT} Security Incidents}
\label{sec:reaction}

To support and prepare actors in the electric power system to react to \ac{ICT} security incidents, MEDIT provides them with both, actionable incident response guidelines and a realistic simulation environment for security training.

\subsection{Actionable Incident Response Guidelines}

Reacting to cyber incidents and \ac{ICT} failures does not only require tools but also operational guidelines with clear recommendations for actions.
However, most guidelines in place (to comply with different security standards, e.g., ISO 27001~\cite{iso-27001} or the German bill on IT security~\cite{IT-Sicherheitsgesetz}) are mainly concerned with organizational measures, specifying who must be informed and what organizational processes should be carried out.
Clear and specific instructions on technical levels are generally not provided.
With MEDIT, we want to fill this gap by providing \emph{actionable incident response guidelines}.

\begin{figure}
\centering
\includegraphics{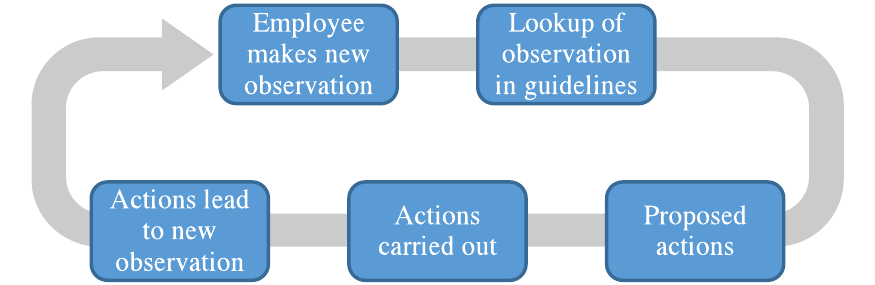}
\caption{Our actionable incident response guidelines support employees in reacting to \ac{ICT} security incidents through an iterative observation/action cycle.}
\label{fig:observation_circle}
\vspace{-1em}
\end{figure}

The goal of our guidelines is to provide distinct and precise actions for all employees who potentially have to deal with a \ac{ICT} failure or cyber attack - especially those with no background in \ac{ICT}, such as control center personnel.
As illustrated in Figure~\ref{fig:observation_circle}, our core idea is to describe observations from employees' point of view.
Based on each observation, different actions are proposed, which then can lead to new observations, providing a starting point for the next iteration.

To demonstrate our approach, we provide a brief example:
We consider an operator, who detects implausible measurement in the \ac{SCADA} system.
The operator uses this initial observation as entry point for the work with our guidelines.
To enable fast lookup of relevant instructions in the guidelines, all observations are categorized and indexed.
In this example, the corresponding proposed action is to try to identify the source of the implausible measurements, enlisting other technicians if necessary.
After performing the necessary steps, the operator (in our example) derives as new observation that one distinct \ac{RTU} is the source of the implausible measurements.

Looking up this new observation in the guidelines, the operator is instructed to inspect this \ac{RTU}'s logfiles and configurations.
In our example, the operator concludes that the \ac{RTU} configuration was altered and unscheduled maintenance access was logged, potentially constituting a cyber security incident.
Consequently, the information security manager and other stakeholders defined in the company's existing incident response guideline have to be informed and the correspondingly defined process has to be triggered.

Currently, our actionable incident response guidelines mainly focus on \acp{DSO} and \acp{VPP}.
However, we are continuously extending these guidelines, by adding new observations, actions, and operators.
The usefulness and practicability of our guidelines is evaluated in different workshops with \acp{DSO}, cyber security experts, and relevant stakeholders.

\subsection{Simulation Environment for Security Training}

Besides having actionable incident response guidelines, those responsible for adequately narrowing down and responding to failures and attacks within the \ac{ICT} infrastructure of power systems require proper training of incident response measures.
To facilitate such security training, we require an environment in which cyber attacks on the \ac{ICT} infrastructure of power systems as well as as corresponding detective and responsive measures can be simulated.

In the past, different frameworks for the simulation of cyber attacks on \ac{ICT} infrastructure have been developed to facilitate research and training in security incident response, attack detection methods, and awareness~\cite{topham_cyber_2016}.
E.g., SECPSIM~\cite{vellaithurai_secpsim_2013} provides a training simulator for power system infrastructure security based on mathematical modeling.
From a different perspective, BREACH~\cite{uetz_breach_2017} simulates a small company network, including legitimate user behavior and cyber attacks against the company, to facilitate incident response research and training.
Contrary, the SWaT Security Showdown~\cite{antonioli_gamifying_2017} is a capture-the-flag-like security training environment realized on top of the MiniCPS simulator for cyber-physical networks.
While providing a general good starting point for realizing a simulation environment for security training for the \ac{ICT} infrastructure of distribution grids, these existing approaches either do not consider the specifics of power systems or lack a realistic simulation of the \ac{ICT} infrastructure.

Thus, MEDIT will provide a simulation environment for security training specifically tailored to the training for reactions to \ac{ICT} security incidents within the \ac{ICT} infrastructure of electrical power systems.
To this end, we couple an existing framework for the simulation of company networks~\cite{uetz_breach_2017} with both an \ac{ICT}/energy co-simulation environment (cf.\ Section~\ref{sec:co-simulation}) and a realistic distribution network laboratory (cf.\ Section~\ref{sec:laboratory}).
To this end, we will connect these networks with virtual network interfaces (TUN/TAP).
We can selectively insert virtual networking components, e.g., firewalls, routers, or \acp{IDS} (cf.\ Section~\ref{sec:intrusion-detection}), thus modeling realistic \ac{ICT} infrastructure (cf.\ Section~\ref{sec:background}).
To connect to real world components, e.g.,  within our distribution grid laboratory (cf.\ Section~\ref{sec:laboratory}), the virtual network interfaces can be bridged to physical interfaces, thus realizing hybrid training environments.
In this setting, we can control which traffic may flow between company and \ac{OT} network, down to individual \acp{RTU}.

This training environment will allow training operators to run a variety of attacks, covering the different attack vectors (cf.\ Section~\ref{sec:background}).
To explore the network or exfiltrate information, they can run network scans or copy network traffic to their own machines.
Other attacks may send rogue control commands to \acp{RTU}, causing them to malfunction or even destabilize the power system.
In the most mischievous case, attacks may manipulate critical components while simultaneously falsifying sensor readings as cover-up.

By coupling the different components of MEDIT, our training environment will provide trainees with a unique opportunity, as we can run attacks that specifically target power systems in an appropriate and realistic environment.

\section{Conclusion}

Fundamental changes in power supply lead to an increased deployment of \ac{ICT} and interdependencies between primary grid operation and \ac{ICT} across all power system actors.
Consequently, \ac{ICT} becomes an increasing target for attacks and failures that can have serious impact on supply security and system stability.
To counter this threat, existing \ac{ICT} security measures must be enhanced with an anomaly monitoring as well as fast and actionable responses to potential incidents.

In this paper, we presented MEDIT, a comprehensive collection of methods to prevent, detect, and react to \ac{ICT} attacks and failures for various actors in electrical power systems.
As a foundation of our work, we reported on our research and validation environment, including an \ac{ICT} energy co-simulation environment as well as a distribution grid laboratory.
These two components can be flexibly coupled to assess the impact of \ac{ICT} attacks and failures as well as to validate the developed methods for prevention, detection, and reaction.
To prevent \ac{ICT} failures and attacks, we presented an approach for an \ac{ICT} monitoring system for distribution grids that allows to observe the status of assets more closely with regard to \ac{ICT} security.
Based on this, our \ac{IDS} for electrical process data envisions to detect anomalies in the area of grid operation as indicators of successful attacks on or failures of IT and \ac{OT} technology.
To support actors in electrical power systems in reacting to \ac{ICT} security incidents, we presented our approach to provide actionable incident response guidelines as well as a realistic training environment for \ac{ICT} security.

Currently, we are working on fully implementing the methods underlying our vision of MEDIT\@.
This includes creating an \ac{ICT} energy co-simulation designed to study \ac{ICT} security, extending our distribution grid laboratory with \ac{ICT}, implementing our methods of \ac{ICT} monitoring for distribution grids and intrusion detection for electrical process data, crafting actionable incident response guidelines, as well as integrating a simulation environment for security training within our research and validation environment.
The developed methods will be validated in simulation and prototypical deployment using our research and validation environment as well as discussed and refined in workshops with \acp{DSO}, cyber security experts, and relevant stakeholders.
Our perspective is to apply our methods and solutions practically in the actual \ac{ICT} infrastructure of the mentioned actors and thus provide important building blocks for securing power systems against the increasing risk of \ac{ICT} attacks and failures.

\noindent\textbf{Acknowledgments.}
This work has received funding from the \ac{BMWi} under project funding reference 0350028.

\end{document}